\documentclass[a4paper]{spie}  

\usepackage{amsmath,amsfonts,amssymb}
\usepackage{graphicx}
\usepackage[colorlinks=true, allcolors=blue]{hyperref}
\usepackage{enumitem}
\usepackage{wrapfig}

\usepackage[square, sort, numbers]{natbib}
\usepackage{journals}

\title{PLC-controlled cryostats for the BlackGEM and MeerLICHT detectors}

\author[a]{Gert Raskin}
\author[a]{Johan Morren}
\author[a]{Wim Pessemier}
\author[b]{Steven Bloemen}
\author[b]{Marc Klein-Wolt}
\author[c]{Ronald Roelfsema}
\author[b]{Paul Groot}
\author[a,b]{Conny Aerts}

\affil[a]{Institute of Astronomy, KU Leuven, Celestijnenlaan 200D, B-3001 Leuven, Belgium}
\affil[b]{Department of Astrophysics, IMAPP, Radboud University, Heyendaalseweg 135, 6525 AJ, Nijmegen, The Netherlands}
\affil[c]{NOVA Optical InfraRed Instrumentation Group, Oude Hoogeveensedijk 4, 7991 PD, Dwingeloo, The Netherlands}

\authorinfo{Corresponding author: Gert Raskin (gert.raskin@ster.kuleuven.be)}

\begin{document} 
\maketitle

\begin{abstract}
BlackGEM is an array of telescopes, currently under development at the Radboud University Nijmegen and at NOVA (Netherlands Research School for Astronomy). It targets the detection of the optical counterparts of gravitational waves.  The first three BlackGEM telescopes are planned to be installed in 2018 at the La Silla observatory (Chile). A single prototype telescope, named MeerLICHT, will already be commissioned early 2017 in Sutherland (South Africa) to provide an optical complement for the MeerKAT radio array. The BlackGEM array consists of, initially, a set of three robotic 65-cm wide-field telescopes. Each telescope is equipped with a single STA1600 CCD detector with 10.5k x 10.5k 9-micron pixels that covers a 2.7 square degrees field of view. The cryostats for housing these detectors are developed and built at the KU Leuven University (Belgium).

The operational model of BlackGEM requires long periods of reliable hands-off operation. Therefore, we designed the cryostats for long vacuum hold time and we make use of a closed-cycle cooling system, based on Polycold PCC Joule-Thomson coolers. A single programmable logic controller (PLC) controls the cryogenic systems of several BlackGEM telescopes simultaneously, resulting in a highly reliable, cost-efficient and maintenance-friendly system. PLC-based cryostat control offers some distinct advantages, especially for a robotic facility. Apart of temperature monitoring and control, the PLC also monitors the vacuum quality, the power supply and the status of the PCC coolers (compressor power consumption and temperature, pressure in the gas lines, etc.).  Furthermore, it provides an alarming system and safe and reproducible procedures for automatic cool down and warm up. The communication between PLC and higher-level software takes place via the OPC-UA protocol, offering a simple to implement, yet very powerful interface. Finally, a touch-panel display on the PLC provides the operator with a user-friendly and robust technical interface. In this contribution, we present the design of the BlackGEM cryostats and of the PLC-based control system.
\end{abstract}


\section{INTRODUCTION}
\label{sec:intro} 
The announcement earlier this year of the first direct detection of gravitational waves by the LIGO detectors brought gravitational wave astrophysics to the centre of science and research. In order to better understand these gravitational wave events and to maximize the science return from their detection, it will be essential to observe them in the electromagnetic domain as well. For that reason, the BlackGEM  project \cite{bloemen2016}  was started by the Department of Astrophysics of the Radboud University Nijmegen (NL), partnering with NOVA (NL) and the Institute of Astronomy of the KU Leuven University (BE). BlackGEM is an array of telescopes dedicated to and optimized for the search of the optical counterparts of gravitational wave events, like the merging of two neutron stars or the collision of a neutron star with a black hole. 

\begin{wrapfigure}{R}{9cm}
\resizebox{9cm}{!}{\includegraphics{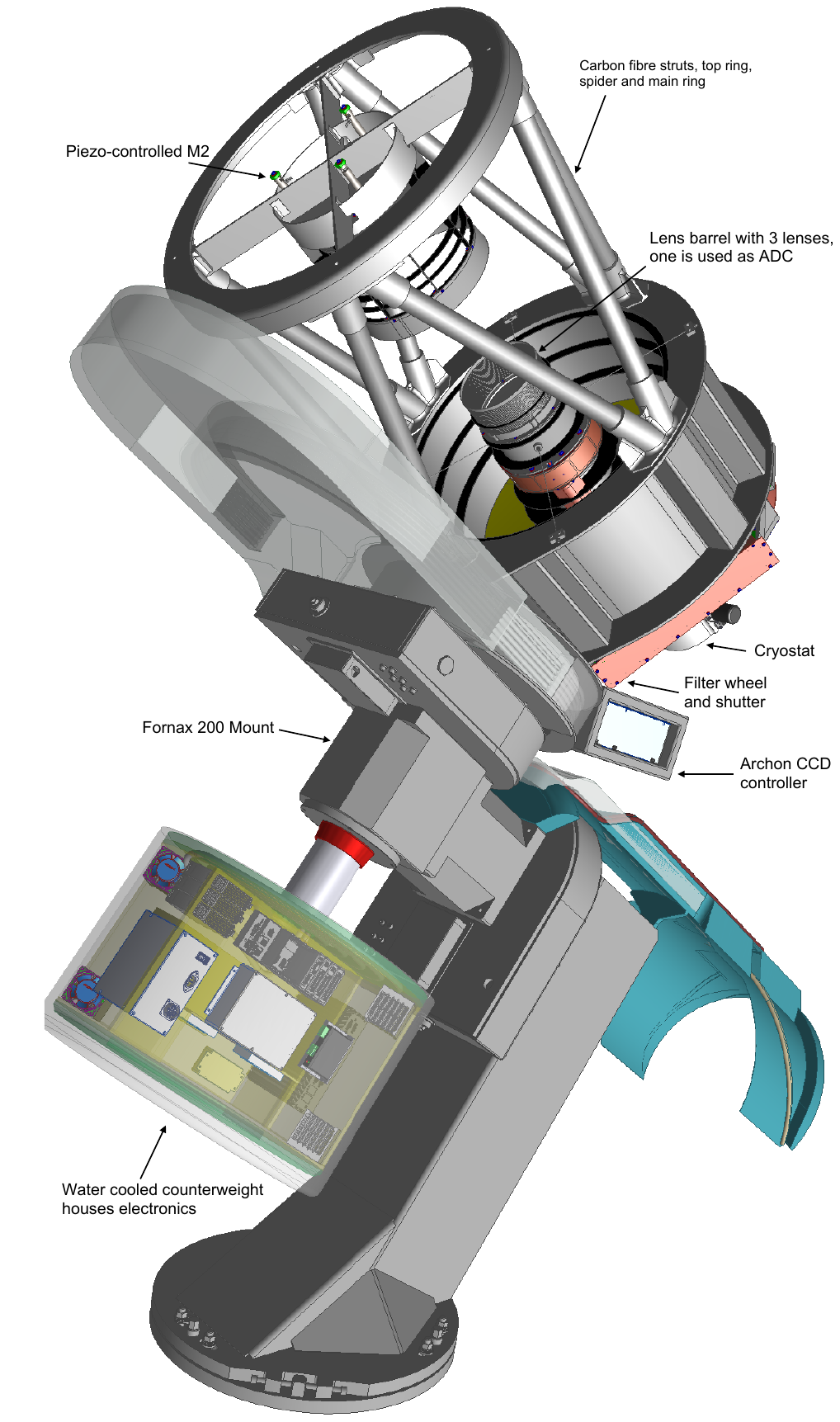}}
\caption{
\label{fig:telescope} 
BlackGEM telescope design.}
\end{wrapfigure}

However, the observations of these electromagnetic counterparts are not straightforward. These sources will probably be faint ($M_{V}\sim22$) and transienting quickly ($\sim$1~day). The large error window of GW detections by facilities like LIGO and VIRGO (typically $\sim$100 square degrees) requires BlackGEM to search with high sensitivity over a very large field of view. Therefore, each telescope will provide seeing-limited imaging over a wide field of 2.7 square degrees. This gives a total sky coverage of 8 square degrees for an array of 3 telescopes.  It is envisioned that in a later stage and with the entrance of new partners in the project, the array will be extended with more telescopes, leading to an even larger total coverage. The array consists of identical Dall-Kirkham Cassegrain telescopes with a 65-cm primary mirror ($f/5.5$) in a carbon fibre structure. The telescopes will be fully robotic with a goal of one-year periods of maintenance-free and unattended operation. Fig.~\ref{fig:telescope} shows a design model of the BlackGEM telescope. We plan to install and commission BlackGEM at the ESO La Silla Observatory in Chile early 2018. 

%

Currently, a prototype for the BlackGEM telescopes is already under construction. After assembly, integration and testing in the Netherlands, this telescope named MeerLICHT, will be installed in Sutherland (South Africa) at the MeerKAT radio array site. There it will complement the radio observations from MeerKAT, the precursor to the Square Kilometre Array (SKA). MeerLICHT will co-point the same field as the radio dishes and provide simultaneous optical observations.

This contribution focuses on the design of the cryostat that houses the CCD detector and the associated cryostat monitoring and control system. The outline of the paper is as follows: in section 2 we introduce the  BlackGEM/MeerLICHT (hereafter solely referred to as BlackGEM) cryostats as well as their detector system, section 3 discusses the thermo-mechanical design of the cryostat, in section 4 we present the PLC-based cryostat control system, followed by a performance discussion in section 5 and some conclusions in section 6.

\section{BlackGEM detector system and cryostat}
\label{sec:cryostat} 

\subsection{Detector system}
\begin{figure}
\begin{center}
\resizebox{\hsize}{!}{\includegraphics{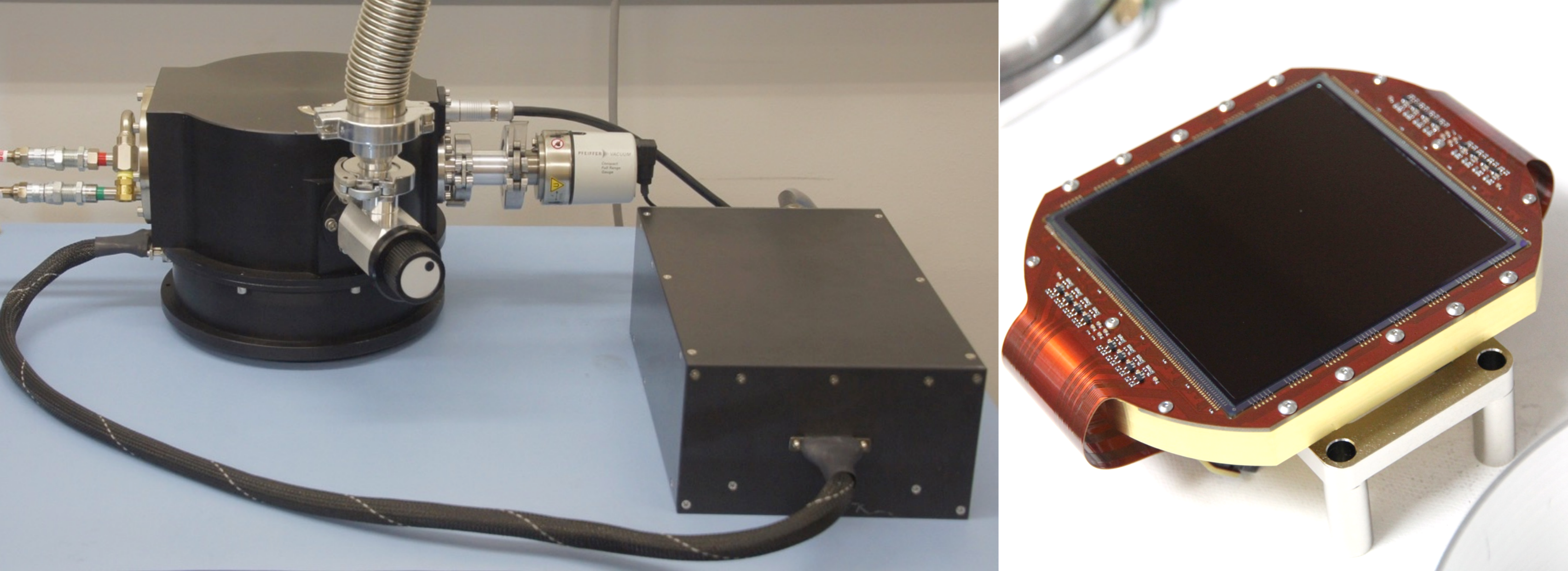}}
\end{center}
\caption
{ \label{fig:ccd} 
Archon CCD controller with BlackGEM cryostat (left) and STA1600 10.5k\,x\,10.5k CCD (right).}
\end{figure}

The BlackGEM detector system is based on the back-illuminated STA1600 CCD from Semiconductor Technology Associates, Inc. (STA), together with their Archon CCD controller. The STA1600 CCD is an impressive piece of silicon (Fig.~\ref{fig:ccd}), consisting of 10\,560\,x\,10\,560 9-$\mu$m pixels. It has an image area of 95\,mm\,x\,95\,mm that corresponds with a field of view of 1.65\,x\,1.65 degrees in the focal plane of a BlackGEM telescope. The spatial sampling amounts to 0.56\,arcsec/pixel. The detector is mounted on a gold-plated Invar base, weighing a solid 1.2\,kg.  

We installed the first STA1600 in the BlackGEM cryostat in April 2016 at STA's facilities in San Juan Capistrano (CA, USA) and performance of the detector proved to be outstanding. Reading out a full frame (110\,MPixels) through the CCD's 16 read ports at 1\,MHz requires only 7\,s.  At values between 5.5 and 6\,e$^-$,  read-out noise is well below specifications for  this speed.
Cosmetic quality and quantum efficiency (surpassing 90\,\% in $g$ and $r$ bands) of this device are excellent too.

\subsection{The BlackGEM cryostat project}
The Institute of Astronomy of the KU Leuven University disposes of  broad experience with the design and construction of cryostat systems for optical detectors, like e.g. the MAIA \cite{raskin2013} and Merope \cite{oestensen10} imagers installed on the Mercator Telescope on La Palma. As the BlackGEM team was initially lacking expertise in this domain, the KU Leuven was invited to join the BlackGEM consortium and take up the responsibility for the cryostat work package.

The BlackGEM telescopes and operational model drove the design of the cryostat. The relatively small telescopes require a compact and low-mass cryostat. Robotic telescope operation has several  implications like remote monitoring and control of the cryostat, high reliability, fail-safe behaviour under all circumstances, a closed-cycle cooling system, etc.  Yearly, a two-week maintenance period is foreseen for the telescopes, implying the need for a vacuum hold time of at least one year. 
As multiple copies will be built, it  should be relatively easy to replicate the cryostat at a reasonable cost.
The operational temperature of the CCD is set at 170\,K with a stability requirement below 0.1\,K.

\section{Thermo-mechanical design}
\label{sec:thermomech}

\subsection{PCC cryostat cooling}
The unmanned operation of BlackGEM excludes the use of liquid nitrogen bath cryostats. 
As the telescopes are small and light weighted, cooling should be vibration free. This directed us towards Joule-Thomson coolers  for the cooling of the CCDs. This type of coolers offers reliable and maintenance-free operation over extended periods of time (more than one year) and is easy to operate remotely. Their only serious disadvantage, especially on a small telescope are the bulky stainless steel tubes between compressor and cryostat, which required us to include a large cable and tube wrap in the telescope mount (see Fig.~\ref{fig:telescope}). 

We selected the Polycold Compact Cooler (PCC, formerly branded with the name  {\it CryoTiger}), equipped with the high-performance cold end for additional cooling power and PT-16 cooling gas. The gas choice was a compromise between cooling power and minimum temperature. The latter is of particular importance for the cryo-pumping efficiency of the charcoal getter and a long vacuum hold time.

\subsection{Cryostat design}
\begin{figure}
\begin{center}
\resizebox{\hsize}{!}{\includegraphics{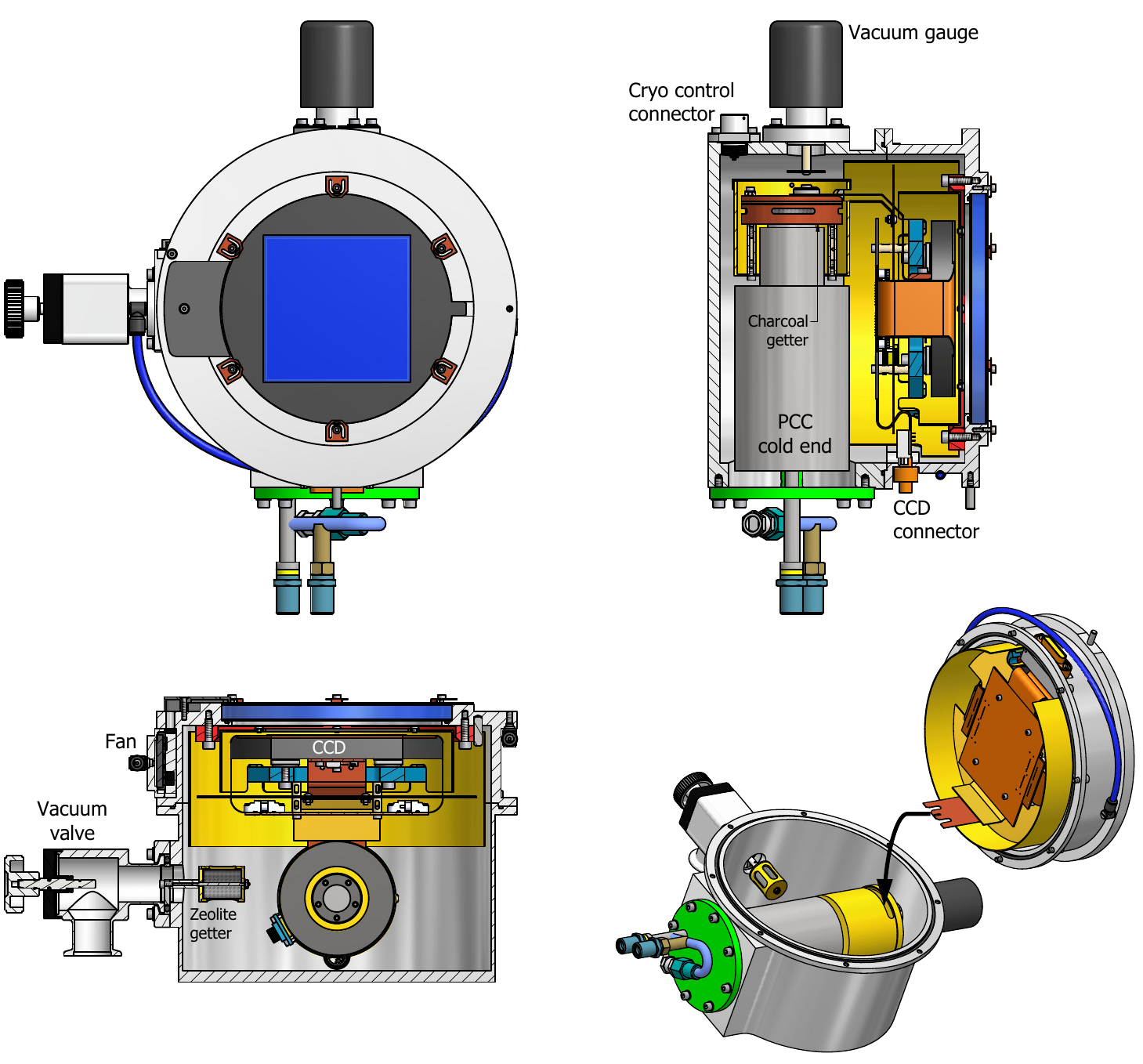}}
\end{center}
\caption 
{ \label{fig:cryostat}
Various views of the cryostat; the bright gold parts are radiation shields surrounding the cold elements.}
\end{figure}

\begin{figure}
\begin{center}
\resizebox{\hsize}{!}{\includegraphics{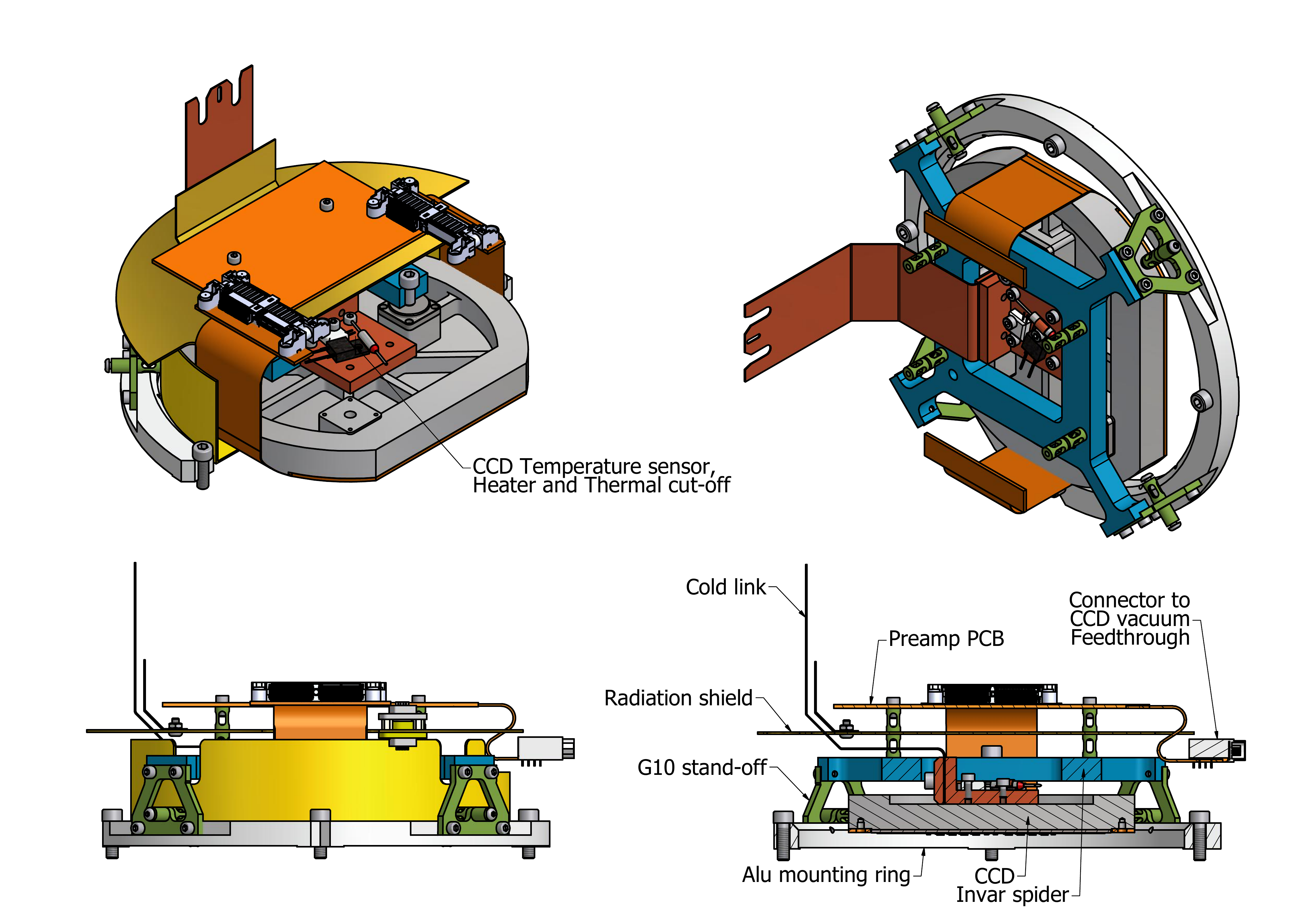}}
\end{center}
\caption 
{ \label{fig:ccdmount}
Various views of the detector mount.}
\end{figure}

Figure~\ref{fig:cryostat} shows the mechanical design of the BlackGEM cryostat. It consists of two main parts: the front element that holds the CCD and the entrance window, and the rear element holding the PCC cold end, vacuum valve and vacuum sensor. Floating low-emissivity heat shields surround all cold parts in the cryostat to reduce radiative heat load. The CCD, packaged on a rigid Invar base, is mounted on a four-arm spider, also made of Invar to ensure homogeneous thermal expansion (figure~\ref{fig:ccdmount}).
This spider is fixed to the front flange through four triangular G10 (fibre glass reinforced epoxy) stand-offs that provide good thermal insulation. While having very high axial and lateral stiffness, the CCD mount still has some radial flexibility for compliance with thermal stress limitations at different temperatures.

The thermal link between detector and cold head consists of a 1-mm thick copper blade. This cold link is connected to the detector through a copper block, fixed at the rear side of the CCD, that also holds the CCD temperature sensor, a heating resistor and a temperature cut-off switch.
The CCD is connected to a printed circuit board (PCB) containing pre-amplifier and differential line-driver circuitry through two flex cables. This PCB is mounted right behind the CCD and connects through a third flex cable to the 100-pins micro-D vacuum feedthrough connector on the cryostat wall.

The PCC cold end is mounted in the rear part of the cryostat in radial direction to limit the overall length of the cryostat. Directly on top of the cold end sits the getter, filled with activated charcoal. The thermal link to the CCD is screw clamped to the getter to ensure good thermal contact.
When opening the cryostat, the cold link needs to be disconnected. To do this, the vacuum gauge is temporarily removed to allow access to the two cold-link clamp screws (Fig.~\ref{fig:cryostat}, bottom right).

\subsection{Vacuum system}
The cryostat was designed with a long vacuum hold time in mind. We target a hold time of at least one year so that vacuum pumping during the yearly telescope maintenance period will be sufficient for maintaining a good vacuum. In order to increase the hold time:
\begin{itemize}[topsep=2pt, itemsep=2pt]
\item[--]
We only used proper and well-cleaned, baked-out materials to limit out gassing.
\item[--]
An oversized getter, filled with 8\,g of activated charcoal, is placed directly on the cooler's cold end to keep its temperature as low as possible as to increase its sorption pumping efficiency. Radiation shields reduce the heat load to help keeping temperature low. 
\item[--]
We limited the number of O-ring (Viton) interfaces (e.g. no separate front or rear flanges: both cryostat parts are machined from solid blocks of aluminium). Diffusion through the O-ring seals is the main cause of vacuum degradation.
\item[--]
A vacuum gauge monitors the pressure in the cryostat. The cold-cathode operation of the gauge provides some additional ion-pumping action in the cryostat, again improving vacuum hold time.
A light baffle (black POM) installed in a custom-made 25KF centring ring blocks the light glow that is emitted by the cathode in the vacuum gauge. 
\end{itemize}

We measured the leak rate of a closed cryostat at ambient temperature: the pressure rises from 4\,x\,10$^{-5}$\,mbar to 3\,x\,10$^{-4}$\,mbar over the course of 8\,hours. For a cryostat volume of 4\,l, this corresponds with a leak rate of less than 4\,x\,10$^{-8}$\,mbar\,l\,s$^{-1}$. We estimate that at a getter temperature of 100\,K only about 1\,g of  activated charcoal, much less than the 8\,g that we have foreseen, is required to adsorb the gasses that permeate into the cryostat during one year of operation.

A second getter, filled with Linde\,5A Zeolite at ambient temperature, traps the water in the cryostat. This is important during warm-up when the water that diffused into the cryostat and was frozen out on the coldest surfaces, will evaporate, potentially leading to condensation and cloud formation. In contrast with the activated charcoal, Zeolite has good capacity to adsorb water at room temperature so that it can keep the warming cryostat dry. The operational disadvantage of  Zeolite is that it needs to be replaced or regenerated at high temperature each time the cryostat is exposed to ambient conditions. This getter is mounted on the centring ring of the vacuum valve, so that it is fast and simple to exchange without having to open the cryostat completely.

\section{Cryostat control}
\label{sec:control}

\subsection{PLC control}
Reliable and safe operation, and extensive monitoring options are essential for a robotic facility. This involves much more than only temperature monitoring and control. It requires safeguarding the CCD under all circumstances, monitoring the vacuum in the cryostat, controlling the PCC compressor, avoiding condensation on the cryostat window, etc. A Programmable Logic Controller (PLC) is the obvious choice for this type of control tasks \cite{pessemier12}.

PLC-based cryostat control not only offers  user-friendly and versatile operation, but it also increases the safety and reliability of the system. This makes it a much more attractive solution, compared to the classical implementation with a dedicated temperature controller (e.g. LakeShore) or using the proper detector read-out electronics to control the temperature of the CCD. This is even more true when several cryostats are installed at one location, all being controlled from a single PLC. As they rely on rugged industrial-type hardware, PLC electronics are robust and well adapted to the environmental conditions at an observatory site. Moreover, PLC software is extremely dependable with fully deterministic execution of the code in real time. In strong contrast with different types of computer hard- and software, PLC manufacturers ensure their user community of long-term ($\sim$ life time of telescope or instrument) support and component availability.

The three BlackGEM cryostats are controlled from a single Beckhoff CP6707 PLC, running TwinCAT software. The CP6707 is a compact touch-panel PC that apart of the PLC program, also runs an HMI (Human Machine Interface) and an OPC-UA communication server. The HMI offers a stand-alone display showing the complete status of the cryostat control system. It  can be used by the operator for configuration (e.g. changing a temperature setpoint, setting the control loop parameters), maintenance and technical interventions (e.g. starting a cool-down or warm-up cycle). The TwinCAT Web interface makes this HMI also available remotely in any web browser.
OPC-UA provides a nearly effortless way to integrate the PLC software in to a higher-level control system. The TwinCAT OPC-UA server exposes the address space of the cryostat PLC and it offers standardized and platform-independent  data access, secure communication, advanced data modelling, alarming and event management, etc. At the client side, only OPC-UA client software is needed for full access to the PLC data. To simplify the development of OPC-UA applications, we developed and use UAF \cite{pessemier12b}, a C++/Python framework that takes care of all OPC-UA technicalities. The PLC software is written in {\it Structured text}, a block-structured high-level language that syntactically resembles {\it Pascal}.

\begin{figure}
\begin{center}
\resizebox{\hsize}{!}{\includegraphics{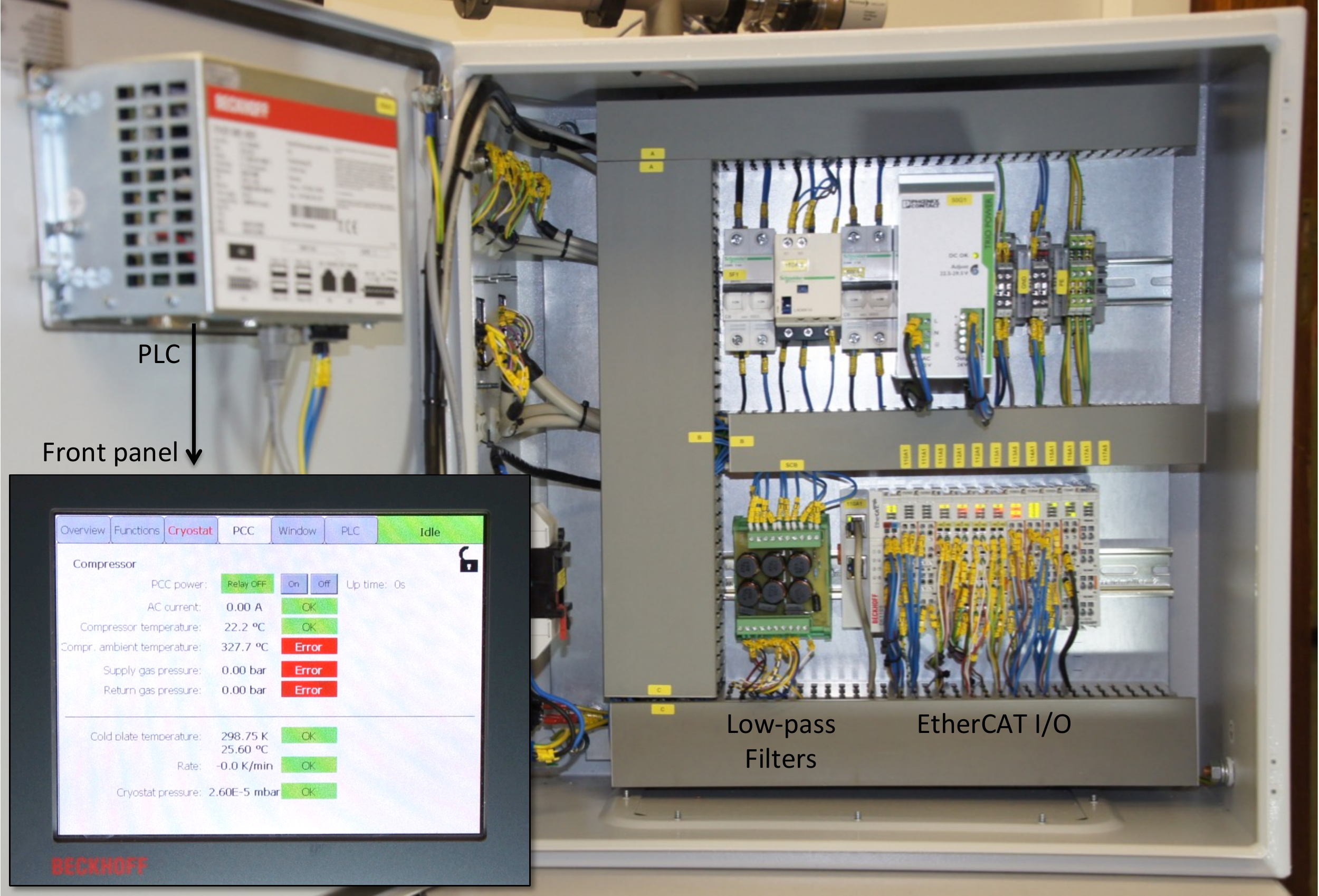}}
\end{center}
\caption 
{ \label{fig:cabinet}
MeerLICHT cryostat control cabinet with the PLC mounted in the cabinet door (top left), the DIN-rail mounted I/O electronics (right) and a front view of the HMI touch-panel display (bottom left inset).
}
\end{figure}

In Fig.~\ref{fig:cabinet}, a picture of the HMI and of the control electronics for the MeerLICHT prototype cryostat are shown. 
The  industrial-style electronics are installed in a rugged electrical cabinet. A similar cabinet will also house the complete control for three or more BlackGEM cryostats, installed at a single site. The PLC is mounted in the cabinet door, making the HMI easily accessible for the operator. The I/O electronics are maintenance-friendly installed on a DIN-rail. They consist of Beckhoff's compact modular I/O terminals that are linked to the PLC through the EtherCAT fieldbus. The system contains modules for temperature measurement, analog input (voltage and 4\,--\,20\,mA), pulse-width modulated (PWM) and digital outputs.

\subsection{Temperature measurement and control}
All temperatures are measured with Pt100 Resistance Temperature Detectors. They have a 4-wire connection to the Beckhoff EL3202-0010 temperature input modules (2 channels per module) with 0.01\,K precision. These modules automatically detect sensor malfunctions such as a broken wire to allow proper alarming.

A resistive heater at the back of the CCD warms up the cooled detector to the required 170\,K. The resistor is driven by the PWM output from a Beckhoff ES2502 module. The PWM output has 16-bit precision and operates at a frequency of 6.67\,KHz. The PWM signal is filtered by an LC low-pass circuit at 100\,Hz. With a 100-$\Omega$ resistor and a 24.5\,V power supply, the maximum heating power amounts to 6\,W. A proportional-integral (PI) control loop with feedback from the CCD temperature sensor stabilises the temperature at 170\,K with a stability well below 0.1\,K. The temperature control loop also monitors the temperature gradient during warm up or cool down. To avoid thermal stress on the detector, the temperature rate is limited to a maximum of 2\,K/minute. 

A similar 6-W resistor heater attached to the getter can warm up the activated charcoal during vacuum pumping in order to increase the getter regeneration speed. It uses the same type of PI loop with a low-pass filtered PWM output to control the getter temperature.
It is worth noting that after LC filtering, a small ripple remains on the PWM signals. This ripple becomes larger for smaller PWM signals. We could not measure any effect of the ripple on the CCD read noise. Neither was the read noise affected by the switched-mode power supply of the PLC electronics, nor by the PCC compressor.

Although the reliability of the PLC-based thermal control system is very high, it could be possible to overheat the CCD in the extreme case of both the CCD and getter heaters run at full power. Therefore,  we added a thermal cutoff in series with each heater that opens the heating circuit above 72\,$^{\circ}$C. For this purpose, we use a Bourns NR72CB0 miniature (1\,x\,2.8\,x\,11\,mm) bimetallic switch. As these very convenient cutoffs are neither specified to be used in vacuum nor at cryogenic temperatures, we have tested their cryogenic performance. Vacuum or thermal cycling to low temperatures does not affect their performance with respect to overtemperature cut-off. They consistently opened at 72\,$\pm$\,3\,$^{\circ}$C. However, in a few cases the cutoff failed to reset to its low-resistance status when the temperature dropped again to the specified 40\,$^{\circ}$C. The cutoff did reset, but the contact resistance increased to several Ohm  after reset instead of returning to normal m$\Omega$ values. This reset behaviour does not really affect their applicability as a protection device against overheating.

\subsection{Remote compressor control}
The remote control of the PCC compressor consists of a simple ON/OFF relay to start/stop the cooler, combined with several monitoring inputs to guard the compressor's condition. We measure the temperature of the compressor ($\sim$65\,$^{\circ}$C during normal operation) and of the ambient air, we monitor the motor's electrical  current consumption (2.5\,--\,3\,A with a short peak to 4\,A at start up for a line voltage of 230\,V). Fig.~\ref{fig:cooldown} shows an example of this monitoring and logging during cool down. Finally, also the pressures in the supply and return gas lines are monitored. The latter gives very valuable information for detecting problems with the PCC system like gas leaks or, more commonly occurring, ice blockage in the cold tip. To protect the cooler against excessive thermal load, starting the compressor is inhibited when the vacuum pressure is above $10^{-3}$\,mbar. For this purpose, we use a Pfeiffer PKR\,251 vacuum sensor that is directly wired to a Bekchoff analog input module, avoiding the need for dedicated sensor interfacing electronics. As soon as the PLC senses the presence of a vacuum sensor, its output is logged and displayed.

\subsection{Avoiding window condensation}
A cold detector will cool the cryostat window by radiation. Thermal modelling shows that without any precautions, radiative cooling can cause a temperature drop of more than 12\,K in the centre of the outside surface of the window. With forced ventilations over the window, this temperature drop can be reduced by a few degrees. We measured the window temperature and found a temperature decrease of $\sim$10\,K.

\begin{figure}
\begin{center}
\resizebox{13cm}{!}{\includegraphics{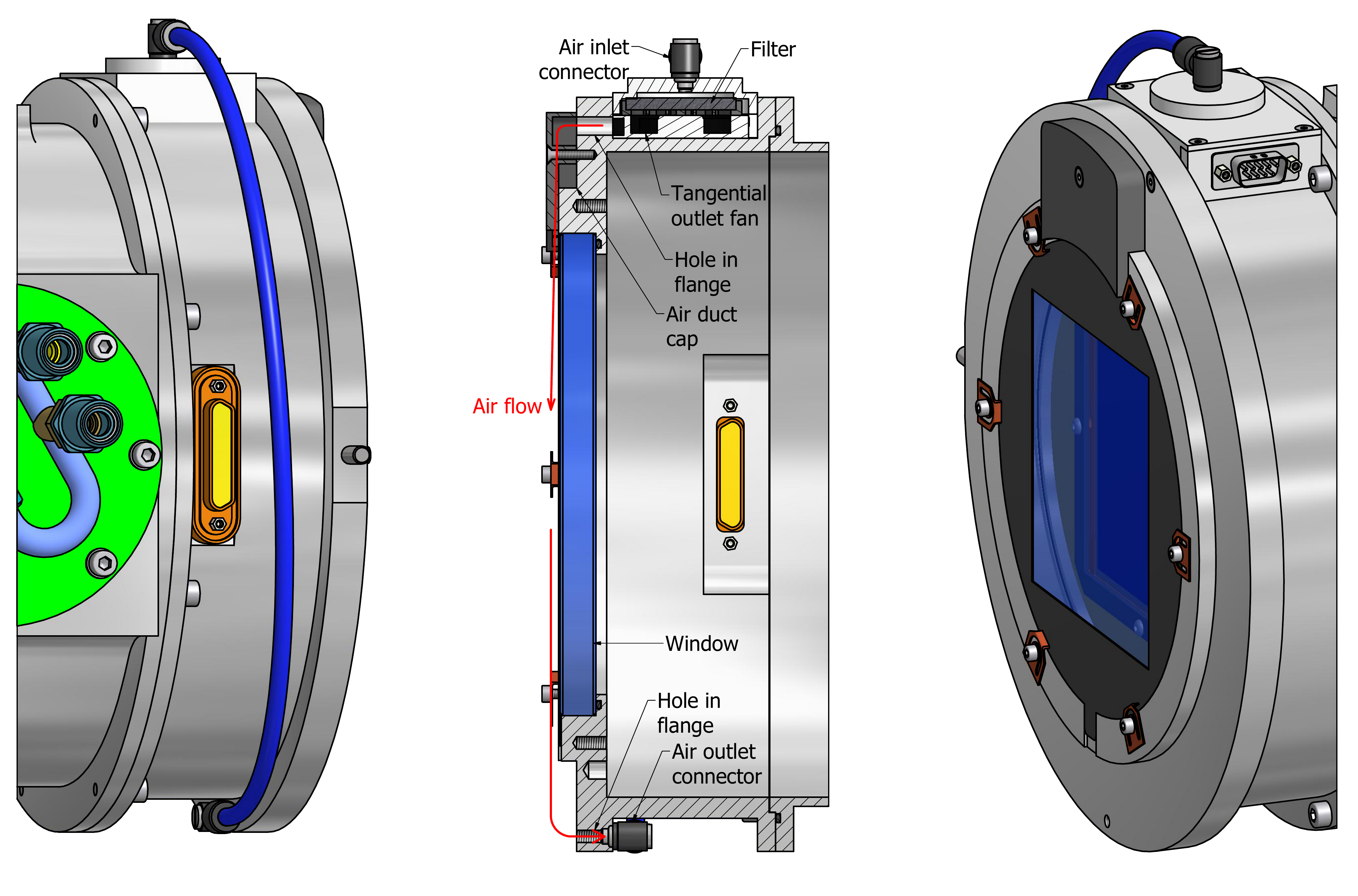}}
\end{center}
\caption 
{ \label{fig:fan}
Fan (top centre) for avoiding condensation on window. The air duct cap directs the filtered airflow over the window, a flexible tube connects the air outlet at the opposite side of the fan to the inlet above the air filter.
}
\end{figure}

In case of high relative humidity (RH), condensation will occur on the outside surface of the cooled window. Under standard conditions and with a RH of $\sim$50\%, the dew point will be 10\,K below ambient or about equal to the window temperature. To avoid condensation on the window, we plan to switch on a small tangential fan that blows air directly over the window as soon as the RH becomes too high (Fig.~\ref{fig:fan}). Furthermore, this airflow can be slightly heated to a few degrees above ambient by a third PWM circuit and a heating resistor in the outlet of the fan (power: $<$\,0.5\,W). The inlet air for the fan is taken from the camera's protected internal environment via a flexible tube and an outlet hole at the opposite side of the window. The circulating air is cleaned by a filter on top of the fan.

\section{Performance}
\label{sec:performance}

\subsection{Cool down}
\begin{figure}
\begin{center}
\resizebox{\hsize}{!}{\includegraphics{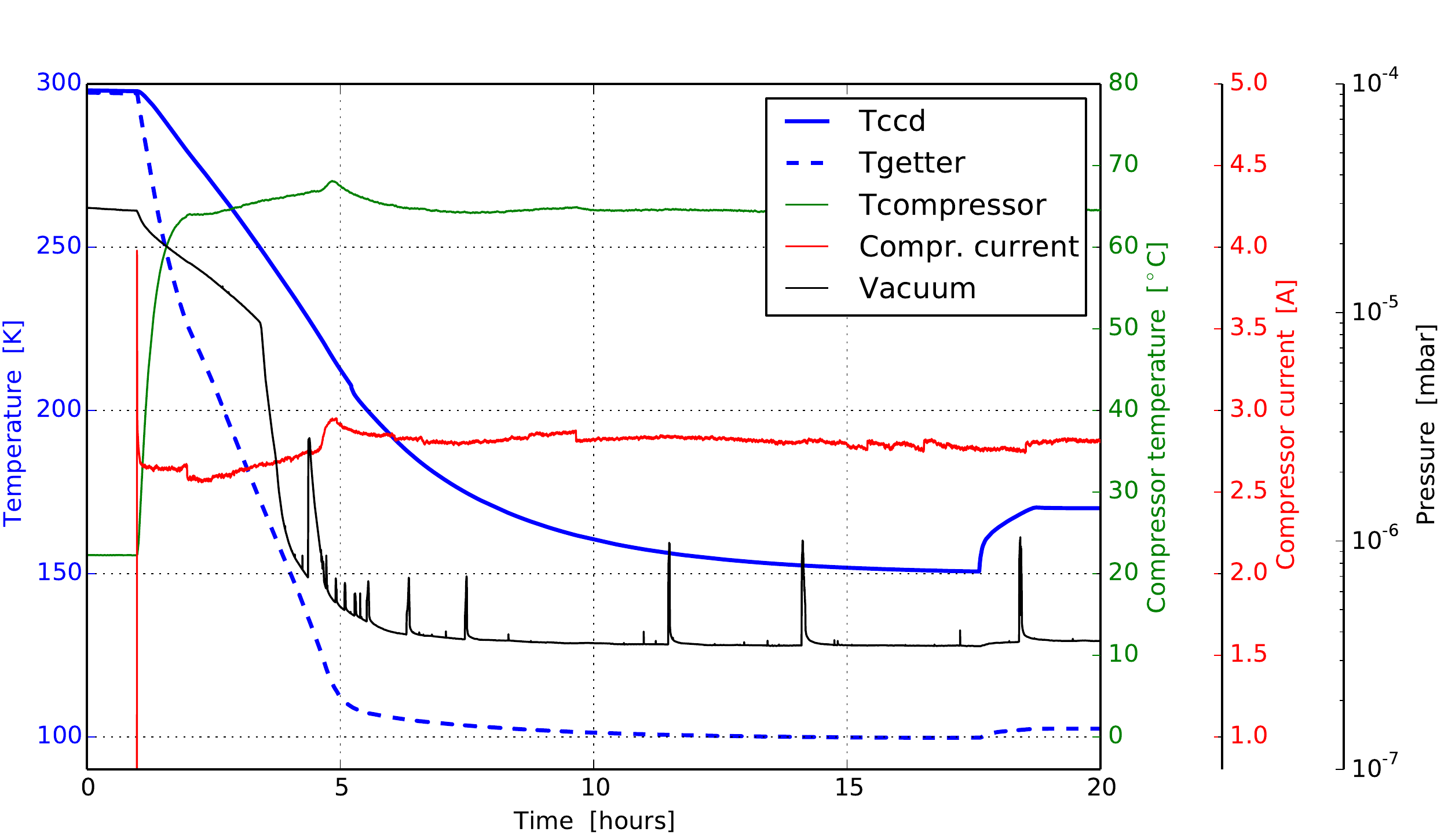}}
\end{center}
\caption 
{ \label{fig:cooldown}
Overview of data recorded during a typical cool-down cycle: detector and getter temperature (blue), compressor temperature (green), compressor current (red) and vacuum pressure (black). The compressor was started at $time\,=\,1$\,h, while the CCD heater was switched on to stabilise the temperature at 170\,K at $time\,=\,17.5$\,h.
}
\end{figure}

In Fig.~\ref{fig:cooldown}, we show a typical cool-down curve for the BlackGM cryostat. The temperature of the getter on top of the cold head drops to almost 100\,K, sufficiently low for efficient adsorption to ensure a long-vacuum hold time. The temperature of the CCD settles at $\sim$150\,K and the heater raises this to 170\,K. The CCD temperature fluctuations over an extended period are smaller than 0.01\,K\,rms. During cool down, the CCD experiences a maximum temperature gradient of 0.5\,K/minute, well below the specified 2\,K/minute. After starting up the cooler, it takes 7\,hours for the CCD to reach the operational temperature of 170\,K. Ambient temperature was 17\,$^{\circ}$C during the recording of these data. Our thermal model of the cryostat shows that the cool-down time increases/decreases by almost one hour when the ambient temperature varies by +/--  10\,$^{\circ}$C.

\subsection{Warm up}
\begin{figure}
\begin{center}
\resizebox{14.5cm}{!}{\includegraphics{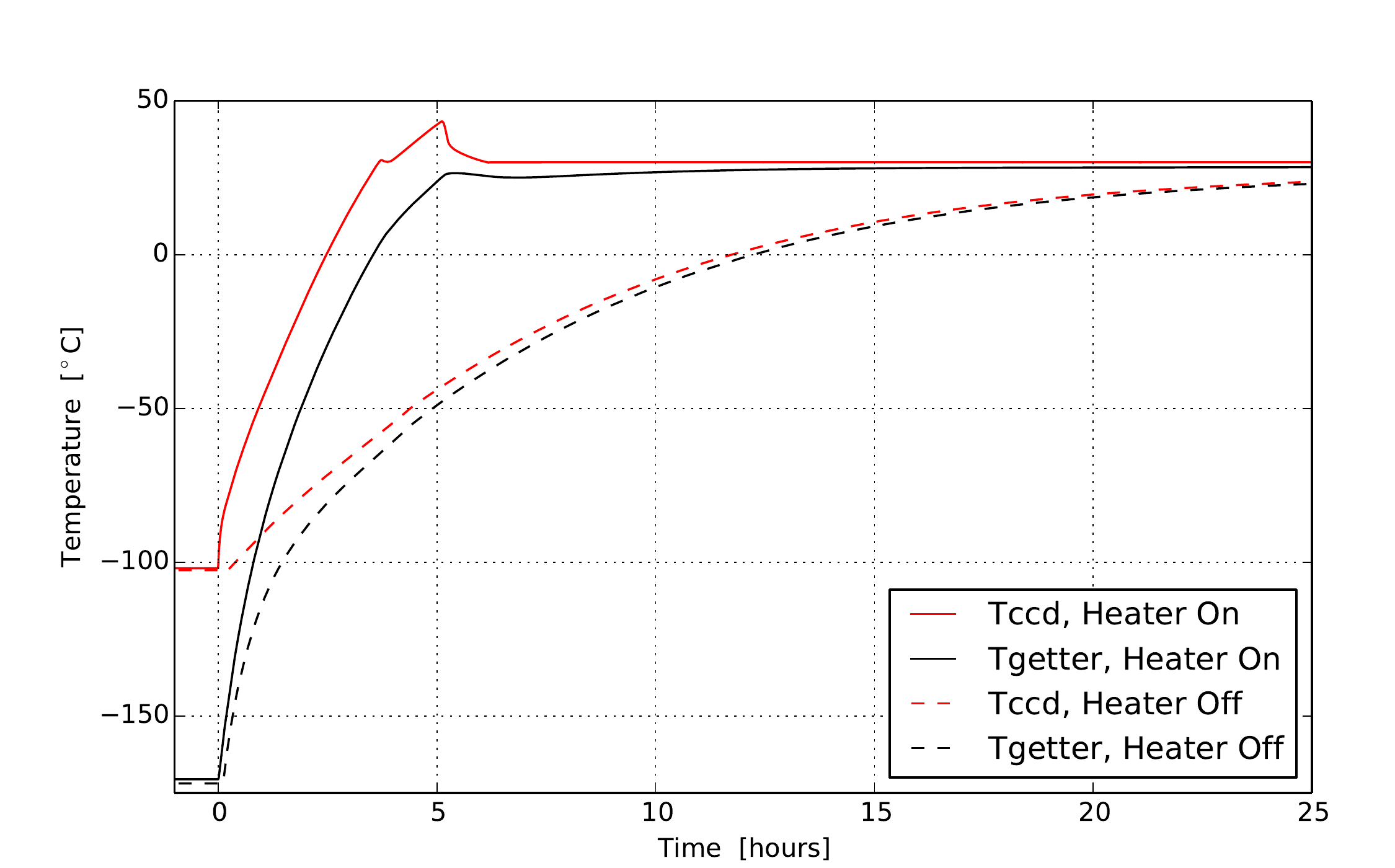}}
\end{center}
\caption 
{ \label{fig:warmup}
Actively controlled (solid curves) and passive (dashed) cryostat warm-up cycle.  
}
\end{figure}

Warming up the cryostat is the most delicate procedure with respect to the detector's integrity. 
It is imperative that the CCD remains warmer than the getter at all times during this process. If this is not the case, gasses that were trapped by the getter may be released and re-condense on the colder CCD, leading to contamination of the imaging surface. Normally, warming up the cryostat will be strictly controlled by the PLC, actively heating up both the CCD and the getter. A temperature difference of at least 20\,$^{\circ}$C between the two will be maintained as long as the temperatures are below zero. This is shown in the solid curves of Fig.~\ref{fig:warmup}. The bump to 40\,$^{\circ}$C in the CCD curve is caused by the getter heater that continues warming up when the CCD heater is already stopped. However, the CCD can easily withstand temperatures above 50\,$^{\circ}$C so this does not pose any risk. The maximum temperature gradient with both heaters at 100\,\% is 1\,K/minute.

In the thermal design of the cryostat, we tried to ensure that also during a passive warm up, i.e. with all active heating switched off as will be the case during a complete power outage, the CCD temperature remains higher than the getter temperature at all times. Although at the start of the  warm up, the CCD is much warmer than the getter, the thermal insulation and thermal inertia of the CCD is also higher. The dashed curves in Fig.~\ref{fig:warmup} show that the CCD temperature indeed stays above the getter temperature. Hence, we do not expect that an uncontrolled warm up will cause CCD contamination. 

\section{Conclusions}
\label{sec:conclusions}
We presented the design of the cryostat and cryostat control for the large BlackGEM detectors.
A prototype has been built and successfully characterised, and will now be installed on BlackGEM's precursor, the MeerLICHT telescope. We argue that PLC technology is very well suited for controlling detector cryostats, a task that requires highly dependable hard- and software. This includes accurate temperature stabilisation as well as robotic control of closed-cycle coolers, and a convenient OPC-UA interface to high-level software.

\acknowledgments 
 
This research was funded by the Big Science program of the Fund for Scientific Research
of Flanders (FWO), by the Department of Physics \& Astronomy of KU\,Leuven and by the 2012 Francqui Prize offered to Conny Aerts.
We thank Kasey Boggs at STA for the constructive collaboration during integration of the CCD in the cryostat.

\bibliography{allreferences}   

\begin{thebibliography}{1}

\bibitem{bloemen2016}
S.~{Bloemen}, P.~J. {Groot}, P.~{Woudt}, M.~{Klein Wolt}, V.~{McBride},
  G.~{Nelemans}, E.~{Koerding}, R.~{Pretorius}, R.~{Roelfsema}, F.~{Bettonvil},
  H.~{Balster}, P.~{Dolron}, A.~{vn Elteren}, E.~{Elswijk}, A.~{Engels},
  M.~{Fokker}, M.~{de Haan}, K.~{Hagoort}, J.~{de Hoog}, R.~{ter Horst},
  G.~{van d Kevie}, S.~{Kozlowski}, J.~{Kragt}, G.~{Lech}, R.~{Le Poole},
  D.~{Lesman}, J.~{Morren}, R.~{Navarro}, W.~{Paalberends}, K.~{Paterson},
  R.~{Pawlaszek}, W.~{Pessemier}, G.~{Raskin}, H.~{Rutten}, B.~{Scheers},
  M.~{Schuil}, and P.~{sybilski}, ``Meerlicht and blackgem: custom-built
  telescopes to detect faint optical transients,'' in {\em Astronomical
  Telescopes and Instrumentation, Society of Photo-Optical Instrumentation
  Engineers (SPIE) Conference Series},  2016.

\bibitem{raskin2013}
G.~{Raskin}, S.~{Bloemen}, J.~{Morren}, J.~{Perez Padilla}, S.~{Prins},
  W.~{Pessemier}, J.~{Vandersteen}, F.~{Merges}, R.~{{\O}stensen}, H.~{Van
  Winckel}, and C.~{Aerts}, ``{MAIA, a three-channel imager for
  asteroseismology: instrument design},'' {\em \aap}~{\bf 559}, p.~A26, Nov.
  2013.

\bibitem{oestensen10}
R.~H. {{\O}stensen}, ``{Observational asteroseismology of hot subdwarf
  stars},'' {\em Astronomische Nachrichten}~{\bf 331}, p.~1026, Dec. 2010.

\bibitem{pessemier12}
W.~Pessemier, G.~Raskin, G.~Deconinck, P.~Saey, and H.~{Van Winckel}, ``Design
  and first commissioning results of plc-based control systems for the mercator
  telescope,'' in {\em Society of Photo-Optical Instrumentation Engineers
  (SPIE) Conference Series},   {\bf 8451}, p.~84512V, Sept. 2012.

\bibitem{pessemier12b}
W.~{Pessemier}, G.~{Deconinck}, G.~{Raskin}, P.~{Saey}, and H.~{Van Winckel},
  ``{UAF: a generic OPC unified architecture framework},'' in {\em Society of
  Photo-Optical Instrumentation Engineers (SPIE) Conference Series},   {\bf
  8451}, p.~84510P, Sept. 2012.

\end{thebibliography}
\bibliographystyle{spiebib} 

\end{document}